\documentclass[letterpaper]{IEEEtran}
\usepackage[svgnames]{xcolor}
\usepackage{tikz}
\usetikzlibrary{positioning,matrix,decorations,decorations.pathmorphing,fit,calc,shapes.geometric,graphs, shapes.misc, datavisualization, datavisualization.formats.functions,intersections, arrows.meta}
\usepackage[utf8]{inputenc}
\usepackage[newfloat, draft]{minted}
\usepackage{csquotes}
\usepackage[colorlinks=true,allcolors=blue]{hyperref}
\usepackage{breakurl}
\usepackage[english]{babel}
\usepackage[T1]{fontenc}
\usepackage[style=ieee,backend=biber]{biblatex}
\usepackage{ifthen}
\usepackage{graphicx}

\graphicspath{{images/}}
\DeclareGraphicsExtensions{.pdf,.jpeg,.png,.jpg,.JPG}

\newcommand{\func}{\textit}

\newboolean{showhints}
\setboolean{showhints}{false}

\newboolean{showtodos}
\setboolean{showtodos}{false}

\ifthenelse{\boolean{showtodos}}{\newcommand{\todo}[1]{{\color{red} TODO:\@#1}}}{\newcommand{\todo}[1]{}}
\ifthenelse{\boolean{showhints}}{\newcommand{\hint}[1]{{\color{blue}XXX:\@#1}}}{\newcommand{\hint}[1]{}}

\bibliography{sources}

\begin{document}

\title{Running Distributed and Dynamic IoT Choreographies}

\author{Jan Seeger\thanks{Jan Seeger is with Corporate Technology, Siemens AG and TU München networking chair}, Rohit Arunrao Deshmukh\thanks{Rohit Deshmuk is with TU Darmstadt}, Arne Bröring\thanks{Arne Bröring is with Corporate Technology, Siemens AG} \thanks{This paper was funded by the EU project 688038, \enquote{Bridging the Interoperability Gap of the Internet of Things}}}

\maketitle

\begin{abstract}
  IoT systems are growing larger and larger and are becoming suitable for basic automation tasks.
  One of the features IoT automation systems can provide is dealing with a dynamic system -- Devices leaving and joining the system during operation.
  Additionally, IoT automation systems operate in a decentralized manner.
  Current commercial automation systems have difficulty providing these features.
  Integrating new devices into an automation system takes manual intervention.
  Additionally, automation systems also require central entities to orchestrate the operation of participants.
  With smarter sensors and actors, we can move control operations into software deployed on a decentralized network of devices, and provide support for dynamic systems.
  In this paper, we present a framework for automation systems that demonstrates these two properties (distributed and dynamic).
  We represent applications as semantically described data flows that are run decentrally on participating devices, and connected at runtime via rules.
  This allows integrating new devices into applications without manual interaction and removes central controllers from the equation.
  This approach provides similar features to current automation systems (central engineering, multiple instantiation of applications), but enables distributed and dynamic operation.
  We also quantitatively evaluate the performance of our chosen approach.
\end{abstract}

\begin{IEEEkeywords}
  Service Composition, Semantic Representation, Internet of Things, Dynamic Service Composition, Service Choreography, Semantic automation
\end{IEEEkeywords}

\IEEEpeerreviewmaketitle

\section{Introduction}\label{sec:introduction}
\todo{Add mention of \enquote{dynamic} functionality}

The Internet of Things (IoT) is making rapid inroads into peoples' everyday life.
The new breed of devices such as the Amazon Echo, Phillips Hue lights and the Nest thermostat allow users to build advanced functionality into their homes.
Using simple configuration tools, users can easily modify their homes and add new devices or reconfigure old ones.

Similar developments are going on in context of automation for commercial buildings.
Today, automation systems for commercial buildings are installed by specialists and typically only configured once during the deployment phase.
Reconfiguring building automation (BA) systems after installation takes a high amount of effort from trained specialists.
Also, the engineering of BA systems is mostly static, adding or removing devices requires the involvement of highly qualified personnel such as electricians and BA specialists for deploying and (re-)~configuring the devices and system.
Expressing dynamic behavior, where devices are added or removed during the operation of the system, is even more difficult.
Sometimes, not even the installation plans of a BA system are available after installation, which means that a \enquote{reverse engineering} of the system becomes necessary for reconfiguration.

Current building automation systems are also centralized to a large degree, with one or more central controllers transferring and converting signals.
The growing computation power of sensors and actuators will make these controllers superfluous, and will allow control algorithms to move into sensors and actuators.
Also, controllers form a single point of failure for the system, where the failure of a controller renders building parts inoperable.
Traditional building automation systems are a form of \emph{orchestration}, where a central controller \emph{orchestrates} the interaction of components.
In this paper, we will move towards a \emph{choreography} of sensors and actors, where the actions of each participant are not controlled by a single controller, but in a distributed fashion.
This transition leads to a number of challenges in management and operation of the system.

We tackle these challenges in this paper by developing a mechanism for the dynamic and automated management of IoT choreographies at runtime.
Our approach is based on semantic technology to describe the structure and configuration of a system based on so-called \emph{Recipes}.
A recipe defines the data flow between IoT devices, so-called \emph{Offerings}, as an abstract template.
We introduce runtime configuration of recipes and allow the definition of communication links to be expressed as rules, so-called \emph{Offering Selection Rules}.
These rules are evaluated at runtime whenever devices are added or removed from the IoT system, in order to keep the recipe choreography running and automatically incorporate new devices.
We illustrate this approach at hand of a use case example from the building domain that is referred to throughout the rest of the paper.
This use case validates the system design and is demonstrates the advantages of dynamic choreographies.

Our use case for evaluation is as follows: A recipe defines the interaction between multiple switches, office lights and motion sensors.
The office lights are controlled by motion sensors and are switched on if motion is detected at any one sensor, but only if any of the switches is enabled.
We will demonstrate in the rest of this paper how our framework allows the centralized creation and decentralized operation of such a system, allowing integration of new devices at runtime.

The remainder of this paper is structured as follows: Section~\ref{sec:background-related} provides background to our work and outlines related work.
In Section~\ref{sec:off--recip}, we describe how services are described in our automation system.
Section~\ref{sec:building-choreography} describes how the selection of service components can be restricted.
In Section~\ref{sec:dynam-chor}, we define the process by which devices are added into the network and how offering selection rules are evaluated to build a choreography.
In Section~\ref{sec:perf-evaluation}, we provide a performance evaluation of the central orchestration component.
We conclude this paper in Section~\ref{sec:conclusions--future} and illustrate future avenues of research.

\section{Background \& Related Work}\label{sec:background-related}

\todo{This section still needs the most work.}

Traditional building automation is based on a static configuration created by highly specialized and developed tools.
The per-application (room lighting, room shading, etc.) \emph{room controller} (RC) provides functionality by accessing connected sensors and actors.
All services available from the controller are preloaded on the controller, and may be parametrized via tools provided by vendors.
When adding new devices, they need to be physically connected to the controller, and the controller needs to be parametrized to use the newly connected devices via the provisioning software (Siemens \enquote{ABT Site}\footnote{\url{http://www.buildingtechnologies.siemens.com/bt/global/en/buildingautomation-hvac/building-automation/building-automation-and-control-system-europe-desigo/room-automation/pages/room-automation.aspx}} tool, or KNX Association \enquote{ETS tool}\footnote{\url{https://www.knx.org/in/software/ets/about/index.php}}).
This configuration process means that room automation functionality is limited to preconfigured functionality on the controller, and that the implementation of dynamic services is difficult.
Additionally, this means that that information on building configuration is only available off-line in the provisioning software configuration file, not on-line in the running system.

Research on building automation tools has led to some advances in the field: Model-based tools have not gained wide acceptance, but represent the current state of the art in building automation system tool research ~\cite{butzin_model_2014}.
Model-based tools allow the configuration and management of BA systems on a higher level.
However, they do not provide the required underlying technology to realize dynamic choreographies as our approach integrating both tools and underlying platform does.
The semantic approach taken by Thuluva et al.~\cite{thuluva_semantic-based_2017} extends automation systems to allow engineering and operation of automation systems.
While these approaches provide functionality for distributed operation of services, no dynamic configuration of the system is supported without user involvement.

In the commercial sphere, \enquote{lightweight} automation services have become popular.
The foremost example here is probably \enquote{If This, Then That} (IFTTT)\footnote{\url{http://ifthisthenthat.com}}, which allows the limited composition of web services and IoT devices with a user-friendly interface.
IFTTT and similar commercial automation services are however limited in their integration with other services.
Integrating external services is difficult due to the inability of these services to export automation descriptions for use in other tools.
By describing services with semantic technology, our system simplifies the creation of external tools to interact with our system.

The \enquote{recipe} concept is a composition language for automation components~\cite{thuluva_recipes_2017}.
Service composition consists of discovering services and connecting them to each other.
In the context of \emph{web} services, there has been intense research activity on composition approaches~\cite{sheng_web_2014}, such as WSDL~\cite{Martin2007} or REST-based techniques~\cite{kopecky2008hrests,verborgh2011efficient}.

Other offerings for service composition are Node-RED\footnote{\url{http://nodered.org}} and FlowHub\footnote{\url{http://flowhub.io}}, which allow the creation of data-flow based compositions.
Flow-based research systems include Calvin~\cite{persson_calvin_2015} and Distributed Node-RED~\cite{giang_developing_2015}.
However, the underlying models used by flow-based composition platforms are not expressive enough to ensure an error-free composition.
The semantic descriptions used in our framework contain enough information to prevent incorrect service compositions.

Apart from the mechanics of composition, there has also been research on dynamically adapting systems.
Using a system of rules allows the automation system to automatically adapt to changing circumstances for autonomous management~\cite{burkert_technical_2015} in a similar vein to our rule-based approach.

\section{Offerings \& Recipes}\label{sec:off--recip}

In the following, we present the recipe and offering models.
These models have been elaborated as part of the BIG IoT project~\cite{broring_enabling_2017} and have been initially introduced in our previous work~\cite{thuluva_recipes_2017}.
Here, we provide an update of these models.
This overview is needed to describe the extensions of these models for dynamic choreographies and making them runtime-ready in the following sections.

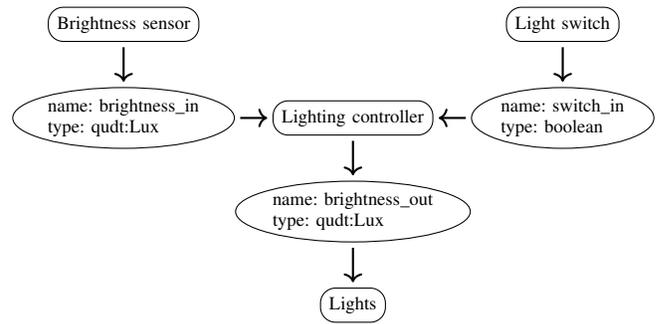
\begin{figure}
  \centering
  \begin{tikzpicture}[
  node distance=6mm and 2mm,
  every node/.style={font=\scriptsize, align=left},
  ingredient/.style={draw,rectangle, rounded corners=2mm},
  input/.style={draw,ellipse, inner sep=.1em},
  every edge/.style={draw,shorten >=2pt,shorten <=2pt,thick},
  interaction/.style={->, draw, shorten >=2pt, shorten <=2pt, very thick}]
  \node[ingredient] at (0,0) (ing1) {Lighting controller};
  \node[input, left=5mm of ing1] (input1) {name: brightness\_in\\type: qudt:Lux};
  \node[input, right=5mm of ing1] (input2) {name: switch\_in\\type: boolean};
  \node[ingredient,above=of input1] (ing2) {Brightness sensor};
  \node[ingredient,above=of input2] (ing3) {Light switch};
  \node[input,below=of ing1] (input3) {name: brightness\_out\\type: qudt:Lux};
  \node[ingredient,below=of input3] (ing4) {Lights};
  
  \draw[->]
  (ing2) edge (input1)
  (input1) edge (ing1)
  (ing3) edge (input2)
  (input2) edge (ing1)
  (ing1) edge (input3)
  (input3) edge (ing4);
  
\end{tikzpicture}
  \caption{\label{fig:light-control-recipe} A lighting control recipe with sensors, switches and lights.}
\end{figure}

\enquote{Recipes} define templates for compositions of \emph{ingredients} and their \emph{interaction}s.
Ingredients are placeholders for \emph{offerings}, devices and services that process and transform data.
Interactions describe the dataflow between these ingredients.
An example recipe is shown in figure~\ref{fig:light-control-recipe} describing a lighting control system.
A lighting controller takes input from brightness sensors, calculates the output brightness through an algorithm (averaging, for example) and outputs the calculated value to the connected lights, but only if one of the switches is switched on.
Inputs and outputs have both a name and a type.
The type is used for matching offerings with ingredients.
This process will be described in Chapter~\ref{sec:building-choreography}.

Offerings describe service or device instances, and how to access these services or devices.
Offerings are specified in a semantic format by the so-called \enquote{offering description}.
Offering descriptions contain information on the in- and outputs of an offering as well as information on how to access the underlying service or device (providing the offering implementation).
An excerpted offering description for our switch-sensor-controller-light example is shown in listing~\ref{lst:ofdesc}.

\begin{listing}[t]
\begin{minted}[fontsize=\scriptsize,numbers=left,numbersep=5pt,xleftmargin=10pt]{json}
{
 "localId": "officeLightOffering",
 "category": "schema:lighting",
 "endpoints": [{
   "uri": 
    "coap://127.0.0.1:5683/LuminaireController",
   "endpointType": "COAP_PUT",
   "acceptType": "APPLICATION_XML",
   "contentType": "APPLICATION_XML"}],
 "requestTemplate": 
  "<dimmableValue>@@brightness@@</dimmableValue>",
 "responseMapping": null,
 "inputData": [{
   "name": "brightness",
   "valueType": "xsd:float"}],
 "outputData": [],
 "extent": {"city": "Munich"}
}
\end{minted}
\caption{\label{lst:ofdesc} Example offering description for a CoAP-enabled office light.}
\end{listing}

The offering description contains functional as well as non-functional properties.
Functional properties describe the implementation of the offering (e.g. which web service endpoint this offering accesses and procotol and payload of the request), while non-functional properties describe installation-specific metadata about the offering (such as the price or location of the offering).
Non-functional and functional properties thus correspond to offering \enquote{interface} and \enquote{implementation}, respectively.
In detail, the offering description contains the following information:

The \texttt{inputData} and \texttt{outputData} (lines 14 and 18) functional properties contain information on the types of input and output that this offering consumes and produces.
They are visible in the recipe in figure~\ref{fig:light-control-recipe} as type annotations on the input and output nodes.
Type annotations are URIs referencing for example a term in the schema.org~\cite{guha_schema._2016} or QUDT~\cite{hodgson2011qudt} ontologies.
Additionally, a category is used to classify the offering, for example, into \enquote{smart building} or \enquote{transportation} categories.
While being useful for users during the creation of recipes, type and category properties are also used in the basic matchmaking algorithm described in the next section.

The internal properties \texttt{endpoints}, \texttt{requestTemplate} and \texttt{responseMapping} (lines 4, 11 and 13, respectively) specify how this offering accesses the underlying service or device.
The endpoint describes the adress under which the web service implementing this offering is reachable.
To define and parse communication payloads, the BIG IoT library can be used~\cite{schmid_architecture_2017}\footnote{Available at \url{https://gitlab.com/BIG-IoT/lib-java}}.
The BIG IoT library allows interpolation of input values into URLs, URL queries and request bodies, while the response can be parsed into output values via a simple parser that can be parametrized per offering description.
Supported protocols for endpoints are HTTP and CoAP, with POST, PUT and GET methods supported for both protocols.
Additionally, the asynchronous OBSERVE option is supported for CoAP.
Payloads can have XML or JSON format.

For example, the offering in figure~\ref{lst:ofdesc} allows dimming a light via an XML payload over CoAP as defined in the \texttt{endpointType} and \texttt{requestTemplate}.
Finally, non-functional properties (\texttt{extent} line 19, in this example) contain information about the offering that support their discovery and selection restriction beyond the basic matching algorithm.
Both algorithms (basic matching on functional properties and advanced matching on non-functional properties) will be described in the next section.

The duality between offerings and ingredients is central to our system: It allows us to utilize a recipe as choreography descriptions independent of concrete implementations.
A recipe is concrete enough so it can be successfully created as a blueprint by users using our tools, but so generic that it can be implemented and run using a wide variety of service implementations without requiring modification of the recipe.

In the next chapter, we describe the process of turning a recipe into a runnable instance.

\section{Instantiating Recipes}\label{sec:building-choreography}

\enquote{Instantiating} a recipe refers to the process of replacing ingredients with offerings, resulting in a recipe that's executable.
A recipe may be instantiated multiple times with different offerings, depending on the requirements.
To instantiate a recipe, suitable offerings are selected by their external properties described in Section~\ref{sec:off--recip}.
Then, extra restrictions called \enquote{offering selection rules} can optionally be applied. Finally, a recipe can be executed as a choreography, as described in Chapter~\ref{sec:dynam-chor}.

The matching algorithm to select suitable offerings works as follows: For each ingredient in the recipe, the database is searched for offerings that can replace this ingredient. Replacement is governed by the following algorithm:

Let $i$ be an ingredient, and $o$ an offering.
We also define $\func{category}()$, $\func{inputs}()$ and $\func{outputs}()$ to access the so-named properties of the offering and ingredient description in Chapter~\ref{sec:off--recip}.

Furthermore, we use the \enquote{subclass of} operator $\sqsubseteq$ to express subclass relations.

$o$ can replace $i$ iff:

\begin{itemize}
\item The category of the offering is a subclass of the category of the ingredient: $\func{category}(o) \sqsubseteq \func{category}(i)$.
\item For each input of the offering, the ingredient has at least one input with the same or subclassed type: $\forall in_o \in \func{inputs}(o): \exists in_i \in \func{inputs}(i): in_i \sqsubseteq in_o$.
\item For each output of the offering, the ingredient has at least one output with the same or superclassed type: $\forall out_o \in \func{outputs}(o): \exists out_i \in \func{outputs}(i): out_o \sqsubseteq out_i$.
\end{itemize}

Note that this allows offerings to have fewer inputs than the ingredient, as well as more outputs.
Superfluous outputs and inputs are ignored.
In order to instantiate a recipe, each of the ingredients in the recipe has to be filled by at least one offering.

This purely type-based matching is very generic, but also rather limited.
Realizing simple use cases such as \enquote{control all lights in room 3 via any switch in the same room} would require defining categories specifically for this application scenario (e.g., defining the category type \enquote{lighting in room 3}).

To address this and to keep recipes generic, we introduce the concept of \enquote{offering selection rules} (OSRs), which allow users to specify offerings that should participate in a recipe in fine-grained detail.
These rules are attached to an ingredient and specify additional requirements on its non-functional properties that an offering needs to provide to be considered for filling this ingredient.

Offering selection rules are evaluated on the non-functional properties of an offering (see Section~\ref{sec:off--recip}).
Non-functional properties can include location of the component, owner of the component or the energy efficiency of this component.
Because the offering description is specified in a semantic format, non-functional properties can be extended easily.

OSRs can query these properties for equality or inequalities to a literal value.
Multiple OSRs can be composed using boolean operators \texttt{AND} and \texttt{OR}.

Additionally, the cardinality of an ingredient can be specified using OSRs.
This means that the minimum and maximum number of offerings replacing an ingredient can limited.

Using this set of OSRs, it is possible to constrain recipes in complicated ways going beyond the basic matching algorithm.
The light recipe from figure~\ref{fig:light-control-recipe} could for example be constrained to only match lights, sensors and switches in room A, with the controller not being constrained to a certain location, but to a cardinality of one.
Instantiating the recipe would then result in one controller being connected to all sensors, all lights and all switches in room A.
By adding a different set of OSRs to the system, the recipe might be constrained to room B.

This functionality is provided in current automation systems by defining templates that describe the a single deployment and then instantiating these templates multiple times, once for each room.
Templates are not held available during the runtime of the system, and thus cannot be reevaluated for dynamic operation of the system.
Using OSRs, the policy that led to the system's current configuration is always accessible and available.
The policy can thus be reevaluated on system changes, something that is not possible with the template-based approach.

\begin{figure}[t]
  \centering
  \begin{tikzpicture}[
  every node/.style={font=\scriptsize, draw, rectangle, rounded corners=2pt},
  edge label/.style={font=\scriptsize,draw=none, auto, xshift=-2pt},
  curved/.style={draw, thick, -Latex[]},
  every edge/.style={draw, thick, -Latex[]}]
  \node (recipe) {Recipe};
  \node[right=of recipe] (rrc) {RRC};
  \node[right=of rrc] (irc) {IRC};
  \node[right=of irc] (osr) {OSR};
  \draw (recipe) edge node[edge label, near end] {*} node[edge label,near start] {1} (rrc)
  (rrc) edge  node[edge label, near end] {*} node[edge label, near start] {1} (irc)
  (irc) edge  node[edge label, near end] {*} node[edge label, near start] {1} (osr);
  \draw (rrc.south east) edge[bend right, shorten >=-1pt,shorten <=-1pt]
  node[edge label,pos=0.9,swap,xshift=0] {1}
  node[edge label,pos=0.1,swap,xshift=0] {*}
  (osr.south west);
\end{tikzpicture}
  \caption{\label{fig:relat-recip-osrs} Relation of Recipes and OSRs}
\end{figure}
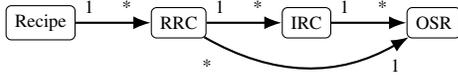

Conceptually, we have implemented these concepts as follows: Recipes, offerings, ingredients and OSRs are stored in an Apache Jena triple store\footnote{\url{https://jena.apache.org/}} as a semantic graph.
The objects in this semantic graphs are \emph{recipes}, \emph{recipe runtime configurations}, \emph{ingredient runtime configurations} and finally \emph{offering selection rules}.
The relations between those concepts are shown in figure~\ref{fig:relat-recip-osrs}.

Recipes are designed using the recipe design tool described by Thuluva et al.~\cite{thuluva_recipes_2017} and are stored in the central repository.
Recipes are then instantiated by creating a \enquote{recipe runtime configuration} (RRC) for this recipe.
Each RRC describes a specific instantiation of a recipe.
A RRC is an installation-specific instantiation of a recipe, because it can (and often will) contain restrictions on non-functional properties (such as the location) of offerings.
Recipes, on the other hand, describe installation-independent patterns of interaction.

The RRC can contain per-recipe OSRs such as cardinality, and always contains per-ingredient information called \enquote{ingredient runtime configuration} (IRC).
Each RRC contains multiple IRCs, one for each ingredient in the recipe.
IRCs contain runtime information describing the current cardinality of the ingredient, as well as the offerings which are currently implementing the ingredient and finally, offering selection rules (OSRs) restricting the set of offerings that can replace this ingredient.

The specification of a service composition as a recipe refined by a set of queries allows the creation of dynamic systems.
We describe the realisation of such systems in the next section.

\section{Design for Enabling IoT Choreographies}\label{sec:dynam-chor}

The concept of OSRs allow the addition of offerings into a running system without manual intervention.

\begin{figure}
  \centering
  \begin{tikzpicture}[
  regular/.style={draw, shorten >=2pt, shorten <=2pt},
  every node/.style={font=\scriptsize},
  interesting/.style={regular,very thick}
  ]
  \node[draw, cylinder,shape border rotate=90, align=center,aspect=0.25] (triplestore) at (-1,0) {Triple \\store};
  \node[draw, rectangle, left=5mm of triplestore,name path=controller path] (controller) {Controller};
  \draw[regular, <->] (controller) -- (triplestore);

  \foreach \x/\dist in {1/-15mm,2/-5mm,3/5mm,4/15mm} {
    \node[draw, rectangle, anchor=east] (device-\x) at ($(controller.west) - (8mm, \dist)$) {Component \x};
  };
  \foreach \x [evaluate=\x as \prev using int(\x -1)] in {3,4} {
    \draw[regular, ->] (controller) -- (device-\x.east);
    \draw[regular, ->] (device-\prev) -- (device-\x);
  };

  \draw[regular, ->] (controller) -- (device-2.east);
  
  \draw[->, interesting, shorten >=2em] ($(device-1.east)!2pt!90:(controller)$) -- ($(controller)!2pt!-90:(device-1.east)$);
  \draw[->, interesting, shorten <=2em, shorten >=4pt]
  ($(controller)!2pt!90:(device-1.east)$) -- ($(device-1.east)!2pt!-90:(controller)$)
  node[midway, right, align=left,anchor=south west]
  {1: Engine registers with OD \\4: Controller sends InDes to engine};

  \draw[->,interesting] (device-1) -- (device-2) node[left, midway, align=left, xshift=-8mm] {5: Component participates\\in recipe};

  \node[below=2mm of controller, anchor=north west, align=left] (label2) {2: Repo finds\\matching ingredients\\3: Repo generates InDes for\\each matching ingredient};
  
\end{tikzpicture}
  \caption{\label{fig:incorp-new-devic} Incorporation of new devices into orchestration}
\end{figure}
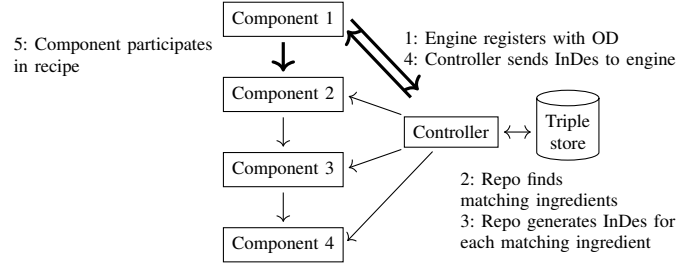

To realize this functionality, our system consists of three parts, as seen in figure~\ref{fig:incorp-new-devic}:

\begin{itemize}
\item A component for computing choreographies (\enquote{controller})
\item A triple store for data storage
\item An \enquote{engine} running on participating components
\end{itemize}

The controller is the central component of our system.
The controller instantiates recipes, and handles the addition and removal of offerings using the triple store in the background for persistent data storage.
The \enquote{engine} implements a gateway and enables devices to participate in the system.
This is done by accepting input from other engines running on other components, passing this input to the offering implementation via the mechanisms described in Section~\ref{sec:off--recip}, parsing the output of the implementation, and sending it to the next offerings currently part of the recipe.
Thus, recipes are turned into distributed choreographies.

In the future, the engine will run on smart sensors and actuators directly, and enable direct integration of these devices.
Currently, the engine is on Raspberry Pi single board computers connected to hardware devices.

It is crucial to note that the controller is \emph{not} a single point of failure.
Without the controller, all recipes will continue operating, only the addition and removal of components is impacted.

The workflow for realizing dynamic choreographies is shown in figure~\ref{fig:incorp-new-devic}.
When a new component is connected to the network, the engine registers at the controller with its offering description (OD) (step 1).
This offering description includes all the information necessary for deciding whether the component should be part of a choreography.
The offering description is added to the triple store.

In step 2, the controller finds all matching IRCs for an offering by computing the matching between the IRC's ingredient and the new offering with the algorithm in Section~\ref{sec:building-choreography}.
Then, for each IRC matching the new offering, all associated OSRs are evaluated.
This is done by serializing the OSRs to SPARQL queries~\cite{harris_sparql_2013} and running them on the triple store.
This results in a number of IRCs that the new component will be added to.

\begin{listing}[t]
\centering
\begin{minted}[fontsize=\scriptsize,numbers=left,numbersep=5pt,xleftmargin=10pt]{json}
{"offering": "bigiot:light-control",
  "recipeRuntimeConfiguration": 
    "bigiot:rrc1",
  "outputs": {
    "brightness": {
      "http://lamp1/input": 
        "on\_off",
      "http://lamp2/input":
        "on\_off"
    }},
  "inputs": {
    "bigiot:sensor1": ["sensorin"],
    "bigiot:switch1": ["switchin"]}}
\end{minted}
\caption{\label{lst:example-inter-descr} Example interaction descriptor for a lighting controller connected to two lights and one sensor and switch.}
\end{listing}
From this information, \emph{interaction descriptors} (InDes) are generated in step 3.
Each interaction descriptor describes the communication behavior of one device as part of a choreography.
InDes' are derived from the recipe by finding all other offerings that an offering should communicate with and accept input from.
An example interaction descriptor is shown in listing~\ref{lst:example-inter-descr}.
An interaction descriptor contains information on where to send the outputs of this offering (lines 5--12) and which inputs to expect (lines 13--16).

Finally, interaction descriptors are sent to each device participating in the choreography (step 4), in order to inform it of its communication partners.

Based on the information contained in the InDes, each component has the knowledge to participate in a choreography (step 5).
Thus, each choreography can now run autonomously, with the new component integrated into it.

This process works analogously for offering removal.
When a device unregisters or fails, its offering description is removed from the triple store.
The controller finds all IRCs that contained this offering and removes it.
Optionally, the removed offering may be replaced by another offering already available in the triple store.

In the next section, we will quantitatively evaluate the computational cost of OSR resolution.

\section{Evaluation}\label{sec:perf-evaluation}
To evaluate the scalability of our implementation, we measured the performance of the controller when adding new components.
The computational factor dominating the addition of new components is the matching and resolution of OSRs.
To find the set of ingredients that the new component can replace, all OSRs need to be evaluated.
Thus, it is expected that the computation time for the addition of new components scales with the number of RRCs in the system.

In order to evaluate this, we measured the time between the addition of a component into the database and the conclusion of OSR computation, when a list of all suitable choreographies is available.
To check the scalability of the controller, we measured performance with an increasing number of RRCs ranging from 7 to 700 using a set of 7 OSRs instantiated $n$ times for $n$ from 1 to 100.
The number of components or recipes in the system does not influence the matching performance, since only RRCs are checked for a match with the new component.

Testing was done on a machine with 8 GB of RAM and a 2.4 GHz i5 mobile Intel processor with 4 threads.

\begin{figure}
  \centering
  \begin{tikzpicture}
    \datavisualization [scientific axes, visualize as smooth line, x axis={label=Number of OSRs}, y axis={label=Time in ms}]
    data [separator=\space] {
      x y
      7 12.9215
      14 15.6139
      21 20.0287
      28 27.2350
      35 31.3954
      42 35.8314
      49 42.5043
      56 46.2951
      63 48.3529
      70 59.2845
      77 69.5230
      84 68.3311
      91 74.4902
      98 84.5599
      105 87.2165
      112 98.2285
      119 107.4427
      126 109.0138
      133 119.6255
      140 122.4260
      147 135.6594
      154 143.8361
      161 159.4636
      168 163.5537
      175 168.0322
      182 181.7566
      189 190.3558
      196 187.4866
      203 207.3416
      210 213.5650
      217 227.9465
      224 230.9730
      231 254.8398
      238 244.1707
      245 281.6536
      252 294.9530
      259 295.0532
      266 312.7786
      273 284.6041
      280 309.7454
      287 323.1019
      294 335.3475
      301 341.3630
      308 358.6607
      315 364.7434
      322 389.7746
      329 372.3820
      336 397.3633
      343 415.7228
      350 452.4144
      357 399.4467
      364 460.6495
      371 470.7582
      378 451.8396
      385 479.2675
      392 490.6598
      399 489.7594
      406 539.3505
      413 528.6679
      420 518.2328
      427 608.9891
      434 593.3963
      441 526.4272
      448 622.3448
      455 598.6910
      462 586.1265
      469 678.3524
      476 637.5745
      483 685.8329
      490 711.6073
      497 619.3662
      504 746.1184
      511 717.8471
      518 698.1797
      525 761.7330
      532 691.2753
      539 802.3693
      546 797.0221
      553 783.4881
      560 780.3362
      567 841.1189
      574 824.4469
      581 875.3933
      588 843.2348
      595 900.6881
      602 868.5855
      609 952.2195
      616 902.3861
      623 957.4786
      630 970.1259
      637 990.6656
      644 994.4331
      651 1026.4359
      658 1102.7518
      665 1008.3464
      672 1133.2290
      679 1091.9723
      686 1133.1709
      693 1151.3241
      700 1175.6033
    };
  \end{tikzpicture}
  \caption{\label{fig:evaluation-plot} Performance evaluation of controller}
\end{figure}
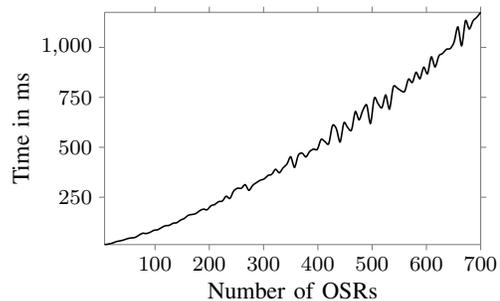

With the controller being the only central component of the system, its performance dominates that of the complete system, and is therefore a suitable indicator of the overall performance.
As can be seen in figure~\ref{fig:evaluation-plot}, the controller scales well enough, with a computation time of one second being broken at about 650 RRCs in the controller.
The system scales quadratically in the number of RRCs, but with low constant factors.
Overall, the controller performance only becomes unacceptable for our purposes with very large systems of more than 650 components.

\section{Conclusions \& Future Work}\label{sec:conclusions--future}

In this paper, we present a concept, implementation and evaluation for running dynamic IoT choreographies.
These dynamic choreographies provide am approach that is novel in IoT environments and particularly useful in the domain of building automation systems.
By expressing service compositions as recipes together with selection rules, IoT components can be dynamically updated and recomposed.
This allows the automatic integration of new components into existing compositions without requiring user interaction.
The choreography approach remove single points of failure, and leverages the computation power of network nodes.
The system is reasonably efficient and scales acceptably with growing numbers of devices.
The quadratic scaling behavior is problematic with very large systems, but performance tuning of the triple store will improve the scaling behavior of the system.
Additionally, the recipe concept is limited in the compositions it can express, only allowing the composition of REST services, without the ability to add custom properties or scripts.
By extending the recipe context in the future, we will be able to express a wider range of automation services, and further reduce the reliance on centralized control algorithms.

Additionally, we are working on using the OSR mechanism as a building block for reliability of orchestrations.
This is a crucial task, since failure of single IoT components may remain unnoticed (or noticed quite late) in distributed workflows.
Using the OSR concept, failures in the orchestration can be automatically corrected if suitable components are available.
This research will make our system ready for more complex deployments.

\printbibliography



\end{document}